\documentclass[journal]{IEEEtran}

\usepackage{cite}

\usepackage{amsmath}
\usepackage{mathrsfs}

\usepackage{algorithmic}

\usepackage{array}

\ifCLASSOPTIONcompsoc
\usepackage[caption=false,font=normalsize,labelfont=sf,textfont=sf]{subfig}
\else
\usepackage[caption=false,font=footnotesize]{subfig}
\fi

\usepackage{url}
\usepackage{graphicx,amsmath,amssymb,amsfonts}
\usepackage{algorithmic,algorithm}

\usepackage{setspace}
\usepackage{tabularx}

\usepackage{xcolor}

\usepackage{bm}
\usepackage{booktabs}
\usepackage{graphicx}
\usepackage{multirow}

\newtheorem{theorem}{\textbf{Theorem}}

\newtheorem{lemma}{\textbf{Lemma}}

\newtheorem{corollary}{\textbf{Corollary}}

\makeatletter

\newcommand{\Rmnum}[1]{\expandafter\@slowromancap\romannumeral #1@}
\makeatother

\usepackage{framed} 

\usepackage{cancel}

\usepackage{flushend}

\usepackage{multirow}

\usepackage{subfig}
\begin{document}
	\bstctlcite{ref:BSTcontrol}
	
	\title{Channel Estimation and Beamforming Design for MF-RIS-Aided Communication Systems}
	
	\author{Zaihao~Pan, Wen~Wang, Gaofeng Nie, Ailing~Zheng, and~Wanli~Ni
		\thanks{The work was supported by Beijing Natural Science Foundation under Grant L232052. Zaihao~Pan, Gaofeng Nie, and Ailing~Zheng are with State Key Laboratory of Networking and Switching Technology, Beijing University of Posts and Telecommunications, Beijing 100876, China (email: \{panzaihao, niegaofeng, ailing.zheng\}@bupt.edu.cn).}
        \thanks{Wen Wang is with the Pervasive Communications Center, Purple Mountain Laboratories, Nanjing 211111, China, and also with the School of Information Science and Engineering, and the National Mobile Communications Research Laboratory, Southeast University, Nanjing 210096, China (email: wangwen@pmlabs.com.cn).}
		\thanks{Wanli Ni is with Department of Electronic Engineering, Tsinghua University, Beijing 100084, China (e-mail: niwanli@tsinghua.edu.cn).}
	}
	
	\maketitle
	
	\newcommand\blfootnote[1]{%
		\begingroup 
		\renewcommand\thefootnote{}\footnote{#1}%
		\addtocounter{footnote}{-1}%
		\endgroup 
	}
	
	\begin{abstract}
		In this letter, we study the beamforming design for channel estimation of multi-functional reconfigurable intelligent surface (MF-RIS)-aided multi-user communications that supports simultaneous signal reflection, refraction, and amplification. A least square (LS) based channel estimator is proposed for MF-RIS by considering both the coupled MF-RIS beams and the introduced thermal noise. With the discrete fourier transform (DFT)-matrix, the MF-RIS beamforming design problem is simplified under the proposed LS channel estimator. The optimal MF-RIS beamforming design that achieves the Cramér-Rao lower bound (CRLB) of channel estimator is obtained with the proposed alternating optimization algorithm.
		Simulation results demonstrate the effectiveness of the proposed beamforming design in reducing the impact of thermal noise.
	\end{abstract}
	\begin{IEEEkeywords}
		Multi-functional RIS, channel estimation, least square estimator, beamforming design.
	\end{IEEEkeywords}
	
	\vspace{-2mm}
	\section{introduction}
    {Reconfigurable intelligent surface (RIS) could enhance the performance of existing systems with a well designed beamformer \cite{2019WUTWC, Sun2024TWCHAP, Sun2024TWCRIS}.}
    A better beamforming design needs a more accurate channel state information (CSI).
    {Since the existing RISs such as passive RIS \cite{2019WUTWC}, active RIS \cite{2021ActiveTWC}, and multi-functional RIS (MF-RIS) \cite{2023IOTJMFRIS} lack integrated signal processing modules \cite{2019WUTWC},}
    the channel estimation between the transmitter and RIS (or between the RIS to receiver) could not be achieved directly.
    Additionally, the large number of elements in the RIS increases the demand for pilots in channel estimation, especially for passive RISs \cite{2022ToCLowcomplexRIS, 2023RISCSI}.
	These factors make accurate RIS channel estimation particularly challenging.
	Most of the existing RIS channel estimation solutions focus on passive RISs-aided scenarios \cite{2020RISChannelEstimationTCOM,2020RISChannelEstimationWCL, 2020ICASSP,2023SRARTCOM}. 
	Specifically, the authors of \cite{2020RISChannelEstimationTCOM} explore the channel estimation in RIS-aided communication systems by regulating the activation states of each RIS element.
    However, this method reduces the accuracy and performance of channel estimation.
    Thus, a discrete fourier transform (DFT)-based RIS channel estimation and reflection beamforming optimization strategy is investigated in \cite{2020RISChannelEstimationWCL}.
    To achieve the optimal RIS beamforming design, the authors of \cite{2020ICASSP} propose the minimum variable unbiased~(MVU) channel estimators, and the optimal channel estimation is derived by reaching the Cramér-Rao lower bound~(CRLB) with RIS phase constraints.
	In addition, in \cite{2023SRARTCOM}, a linear minimum mean square error (LMMSE) estimation method is proposed for the intelligent omi-surface.
	For the system with built-in active components, \cite{2023WCLActiveRISesti} proposes a least square (LS)-based channel estimation scheme for an active RIS-aided system.
    {The channel estimation errors minimization problem is transformed into optimizing the reflective and refractive parameters.
	However, with the function of signal amplification, reflection, and refraction, the channel estimation in MF-RIS is further complicated by the introduced thermal noise and the coupled reflective and refractive beams. 
	Hence the above channel estimation work is unable to employ in MF-RIS communication directly.
    Moreover, numerous studies have verified that MF-RIS can reduce the number of elements while maintaining performance, thereby minimizing the need for pilot signals \cite{2023IOTJMFRIS, 2024TVTMFRIS, 2023WCLMFRIS, 2024CLMFRISUL, 10718320, Sun2024JSACMFRIS}.
	

	To address the issues of the MF-RIS channel estimation, we propose a channel estimator for the MF-RIS communications and jointly design the coupled MF-RIS beams with an {alternating optimization}~(AO) based algorithm.
	Our main contributions are summarized as follows:
	\begin{enumerate}
		\item {In the considered MF-RIS-aided multi-user communication system, we propose a channel estimation technique based on the LS estimator and design the MF-RIS beamforming with the aim of minimizing the estimation error.}
		\item  {Leveraging the property of the DFT matrix, we simplify the MF-RIS beamforming design problem by reaching the CRLB and optimizing the reflective and refractive parameters. To tackle the coupled MF-RIS beams and reduce the thermal noise, we propose an AO-based algorithm, where we derive closed-form expressions for the channel estimation errors and MF-RIS beamforming.}
		\item {Simulation results demonstrate the effectiveness of proposed channel estimators and the proposed beamforming strategies under varying system setups. }
	\end{enumerate}
 
    \vspace{-4mm}
	\section{System Model}
    As shown in Fig. \ref{system_model}, we consider an MF-RIS-aided uplink communication system consisting of an $M$-antenna base station~(BS), an $N$-element MF-RIS, and
    $K$ single-antenna users. Let \( \mathcal{I} = \{T, R\} \) denote the spatial set, with \( \mathcal{K}_R \) (\( \mathcal{K}_T \)) representing the sets of users located in the reflective (refractive) regions. When \( k \in \mathcal{K}_R \), \( i = R \); otherwise, \( i = T \). For the considered system, the pilot length must satisfy \( L \geq N + 1 \) and \( L \geq K \) to ensure unique channel estimation.
	\begin{figure}[t]
		\centering
		\includegraphics[width=0.5 \textwidth]{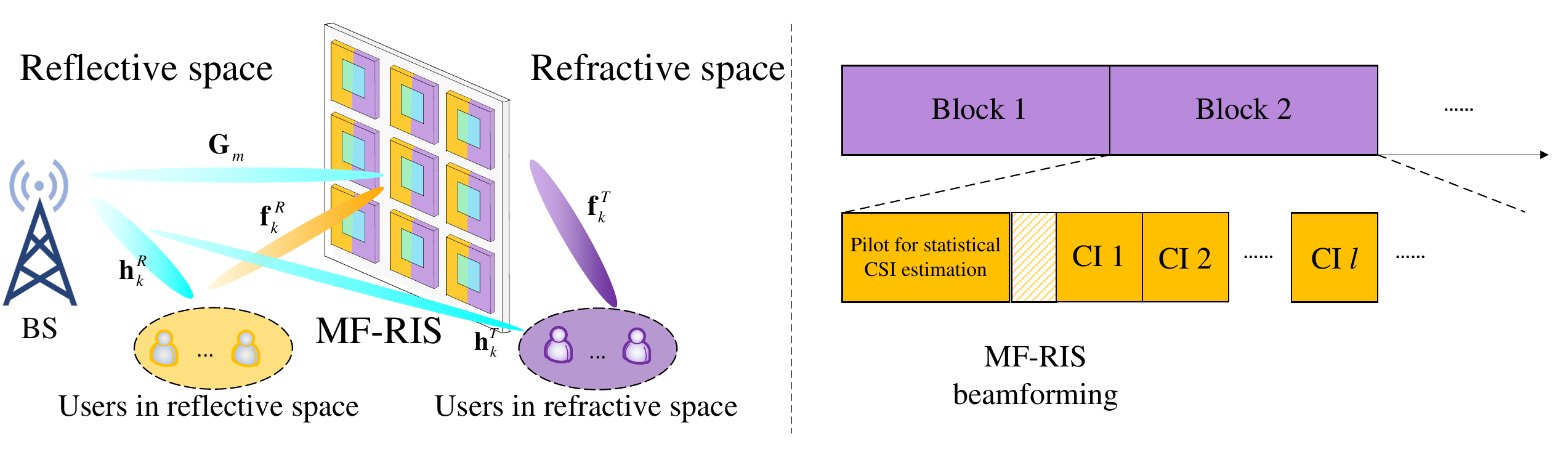}
		\caption{Illustrations of an MF-RIS-aided system and the frame structure.}
		\label{system_model}
	\end{figure}
	\vspace{-2mm}
	\subsection{Channel and Signal Models}
	As shown in Fig. \ref{system_model}, we adopt the block fading model similar to \cite{blockCHRIS}, where the statistical CSI remains constant over multiple coherent intervals~(CI).
	Given the slow variation of statistical CSI, we can reasonably assume stationarity across consecutive block.
	Consequently, the acquisition of CSI is simplified in statistical CSI-based MF-RIS beamforming design. \par
	We denote $\mathbf{h}^i_{k}(l) = [h^i_{1,k}(l), \cdots, h^i_{M,k}(l)]^T \in \mathbb{C}^{M \times 1}, \forall i \in \mathcal{I}, $ as the channel from user-$k$ to the BS at the $l$-th pilot slot, $\forall l \in \{1, 2, \cdots, L\}$, where $[\cdot]^T$ represents the transpose of matrix $[\cdot]$.
    The channels from user-$k$ to the MF-RIS and from the MF-RIS to the $m$-th antenna of the BS in each pilot slot are expressed as $\mathbf{f}^i_k(l) \in \mathbb{C}^{N \times 1}$ and $\mathbf{g}_{m}(l) \in \mathbb{C}^{N \times 1}$, respectively.
    Due to the slow varies assumption, we can simplify $\mathbf{h}^i_{k}(l), \mathbf{f}^i_k(l)$, and $\mathbf{g}_{m}(l)$ as $\mathbf{h}^i_{k}, \mathbf{f}^i_k$, and $\mathbf{g}_{m}$.
    And we assume $\mathbf{g}_{m}$ as the line-of-sight~(LoS) channel.
    We define the reflection and refraction coefficient vectors of MF-RIS as $\bm{\varphi}_{R} = [\varphi_{R,1}, \cdots, \varphi_{R,N}]$ and $\bm{\varphi}_{T} = [\varphi_{T,1}, \cdots, \varphi_{T,N}]$, respectively, where $\varphi_{i,n} = \sqrt{\beta_{i,n}}e^{j\theta_{i,n}}$, $i \in \mathcal{I}$, $\beta_{i,n} \in [0, \beta_{\max}]$, and $\theta_{i,n} \in [0, 2\pi)$.
	For each MF-RIS element, $\beta_{\max} \geq 1$ represents the maximum amplification factor.
	We assume that the reflection and refraction beamforming are independent in each element \cite{2024CLMFRISUL}
    and the MF-RIS beams are stationary in a block. To reduce the pilot contamination between users and the BS, we deploy $K$ orthogonal pilot sequences denote by $\mathbf{s}_k = [s_k(1), \cdots, s_k(L)]^{\rm T}$ and $\lvert s_k(l) \rvert^2 = 1, \forall k$.  
    At the $l$-th pilot slot, all users transmit the pilot symbols simultaneously.
    Define $\mathbf{G}_{m} = \mathrm{diag}\{\mathbf{g}_{m}^{\mathrm H}\}$, in each pilot slot, the signal received at the {$m$}-th antenna is given by
    \begin{equation}
	\setlength\abovedisplayskip{4pt}
	\setlength\belowdisplayskip{4pt}
	\label{origin_signal_model}
	\begin{aligned}
	    y_{m}(l) = &\sum_{i \in \mathcal{I}} \sum_{k \in \mathcal{K}_i} \left[ \sqrt{P_k} \left(h^i_{m,k} + \bm{\varphi}_{i}^{\rm H}\mathbf{G}_{m} \mathbf{f}^i_k \right) s_k(l) \right]  \\
        &+  \bm{\varphi}_{i}^{\rm H}\mathbf{G}_{m} \mathbf{z}_i(l) 
        + n_{m}(l), \forall l,
	\end{aligned}
    \end{equation}
    where ${P_k}$ is the transmit power for user-$k$
    and $\mathbf{z}_i(l) \sim \mathcal{CN}(\mathbf{0}_N, \sigma_{s}^2\mathbf{I}_N)$ and $n_{m}(l) \sim \mathcal{CN}(0, \sigma^2)$ denote the thermal noise at the MF-RIS with the power $\sigma_{s}^2$ and the receive noise in $m$-th antenna with power $\sigma^2$ at the $l$-th pilot slot, respectively. 
    Let $\mathbf{y}_{m} = [y_{m}(1), \cdots, y_{m}(L)]$, then we can stack the signal received at {$m$}-th antenna as 
    \begin{equation}
        \setlength\abovedisplayskip{2pt}
    	\setlength\belowdisplayskip{2pt}
    	\label{signal_model_all_pliot}
    	\mathbf{y}_{m}  =   \sum_{i \in \mathcal{I}}\! \sum_{k \in \mathcal{K}_i} \left[ \sqrt{P_k}\mathbf{\Theta}_{m,i}\mathbf{h}^i_{m,k}\mathbf{s}_k\right] + \bar{\mathbf{n}}_{m},
    \end{equation}
        where $\mathbf{h}^i_{m,k} = [h^i_{m,k}, (\mathbf{f}^{i}_k)^{\rm T}]^{\rm T}$ is the channel vector. We have $\bar{\mathbf{n}}_{m} = \mathbf{\Psi}_{m,R}\mathbf{Z}_R \!+ \!  \mathbf{\Psi}_{m,T}\mathbf{Z}_T \! + \!\! \mathbf{n}_{m} $, where $\mathbf{n}_{m} = [n_{m}(1), \cdots, n_{m}(L)]^{\rm T}$, $\mathbf{Z}_i = [\mathbf{z}_i^{\rm T}(1), \cdots , \mathbf{z}_i^{\rm T}(L)]^{\rm T}$ and
		\begin{equation}
			\mathbf{\Theta}_{m,i}  = 
			\begin{bmatrix} 
				1 & \bm{\varphi}_{i}^{\rm H}\mathbf{G}_{m} \\ 
				\vdots &  \vdots \\
				1 & \bm{\varphi}_{i}^{\rm H}\mathbf{G}_{m}  \ 
			\end{bmatrix} \! , \!
			\mathbf{\Psi}_{m,i}  = 
			\begin{bmatrix} 
				\bm{\varphi}_{i}^{\rm H}\mathbf{G}_{m} & & \mathbf{0} \\ 
				&  \ddots & \\
				\mathbf{0} & & \bm{\varphi}_{i}^{\rm H}\mathbf{G}_{m}  \ 
			\end{bmatrix}.
			\nonumber
		\end{equation}
    After receiving ${y_m}(l), \forall m$ across $l = [1, 2, \cdots, L]$, the BS estimates $\mathbf{h}^i_{m,k}$ and $\mathbf{f}^i_k$ with the conjugate transpose of the user-$k$ pilot sequence, i.e.,$[s_k^{*}(1), \cdots, s_k^{*}(L)]$.
    \begin{equation}
    \setlength\abovedisplayskip{2pt}
	\setlength\belowdisplayskip{2pt}
	\label{signal_model_all_pliot}
	\mathbf{y}_{m,k} \! =  \! \sqrt{P_k} \mathbf{\Theta}_{m,i}\mathbf{h}^i_{m,k} + \mathbf{\Psi}_{m,R}\mathbf{Z}^{'}_R + \mathbf{\Psi}_{m,T}\mathbf{Z}^{'}_T + \mathbf{n}^{'}_{m}, 
    \end{equation} 
    where $\mathbf{n}^{'}_{m} = [n^{'}_{m}(1), \cdots, n^{'}_{m}(L)]^{\rm T}$, $\mathbf{Z}^{'}_i = [\mathbf{z}_i^{' \rm T}(1), \cdots , \mathbf{z}_i^{'\rm T}(L)]^{\rm T}$. Here $\mathbf{z}^{'}_i(l) = \mathbf{z}_i(l)s_k^{*}(l) \sim \mathcal{CN}( \mathbf{0}_N, \sigma_{s}^2\mathbf{I}_N)$ and $ n^{'}_{m}(l) = n_{m}(l)s_k^{*}(l) \sim \mathcal{CN}(0, \sigma^2)$. 
    \par
	The BS estimates $\mathbf{h}^i_{m,k}$ based on the observation matrix $\sqrt{P_k}\mathbf{\Theta}_{m,i}$ with the equivalent noise as $\bar{\mathbf{n}}_{m} = \mathbf{\Psi}_{m,R}\mathbf{Z}^{'}_R + \mathbf{\Psi}_{m,T}\mathbf{Z}^{'}_T + \mathbf{n}^{'}_{m}$, where $\bar{\mathbf{n}}_{m} \sim \mathcal{CN}\{\mathbf{0}_L, \mathbf{C}_{z,m}  \}$ and 
    \begin{equation}
        \setlength\abovedisplayskip{2pt}
        \setlength\belowdisplayskip{2pt}
        \label{equivalent noise}
	\mathbf{C}_{z,m} = \sigma_{s}^2 \left[\mathbf{\Psi}_{m,R}\mathbf{\Psi}_{m,R}^{\rm H} +  \mathbf{\Psi}_{m,T}\mathbf{\Psi}_{m,T}^{\rm H} \right] + \sigma^2\mathbf{I}_L.
    \end{equation}
    \vspace{-8mm}
    \subsection{LS-Based Estimators}
    Note that \eqref{signal_model_all_pliot} is linear for $\mathbf{h}^i_{m,k}$, allowing us to derive a LS estimator for $\mathbf{h}^i_{m,k}$, which also serves as a MVU estimator. 
    $\mathbf{h}^i_{m,k}$ can be estimated as
    \begin{equation}
        \setlength\abovedisplayskip{3pt}
        \setlength\belowdisplayskip{3pt}
        \label{LS_estimation_channel}
        \hat{\mathbf{h}}^i_{m,k} = [\hat{h}^i_{m,k}, (\hat{\mathbf{f}}^i_{m,k})^{\rm T}]^{\rm T} = \mathbf{E}_{m,k}\mathbf{y}_{m,k},
    \end{equation}
    where $\mathbf{E}_{m,k}$ is the LS estimator for $\mathbf{h}^i_{m,k}$, with
    \begin{equation}
        \setlength\abovedisplayskip{3pt}
        \setlength\belowdisplayskip{3pt}
        \label{LS_estimator}
        \mathbf{E}_{m,k}  = \frac{1}{\sqrt{P_k}}(\mathbf{\Theta}_{m,i}^{\rm H} \mathbf{C}_{z,m}^{-1}\mathbf{\Theta}_{m,i})^{-1}\mathbf{\Theta}_{m,i}^{\rm H}\mathbf{C}^{-1}_{z,m}.
    \end{equation}
    Then, the correlation matrix of $\hat{\mathbf{h}}^i_{m,k}$ is calculated by
    \begin{equation}
        \setlength\abovedisplayskip{3pt}
        \setlength\belowdisplayskip{3pt}
        \label{estimation_channel_correlation_matrix}
        \mathbf{C}_{h,m,k} = \frac{1}{P_k} (\mathbf{\Theta}_{m,i}^{\rm H} \mathbf{C}_{z,m}^{-1} \mathbf{\Theta}_{m,i})^{-1}.
    \end{equation}
    The BS can estimate the channel in \eqref{LS_estimation_channel} using each antenna.
    Then, the estimated MF-RIS-BS channel is given by $\hat{\mathbf{h}}^i_{m,k} = [\hat{h}^i_{m,k}, \cdots, \hat{h}^i_{m,k}]$,
    while the user-MF-RIS channels can be estimated with $M$ samples.
    Thus, to reduce estimation errors, we utilize the sample mean of $M$ samples to express the estimation of user-MF-RIS channels, given by
    \begin{equation}
        \setlength\abovedisplayskip{4pt}
        \setlength\belowdisplayskip{4pt}
        \label{f_sample_mean}
        \hat{\mathbf{f}}^i_{k} = \frac{1}{M}\sum_{m=1}^{M}\hat{\mathbf{f}}^i_{m,k}, \forall k.
    \end{equation}
    As a result, the estimation error of the considered MF-RIS-aided communication system is expressed as
    \begin{equation}
    \label{estimation_variable}
        \begin{aligned}
            \epsilon &= \sum_{i \in \mathcal{I}} \sum_{k \in \mathcal{K}_i} \mathbb{E}\left[\lVert \mathbf{h}^i_{m,k} - \hat{\mathbf{h}}^i_{m,k}\rVert^2 \right]  \\
            \vspace{-2mm}
            &= \sum_{i \in \mathcal{I}} \sum_{k \in \mathcal{K}_i} \mathbb{E}\left[\lVert \mathbf{h}^i_{k} - \hat{\mathbf{h}}^i_{k}\rVert^2 \right] + \mathbb{E}\left[\lVert \mathbf{f}^i_{k} - \hat{\mathbf{f}}^i_{k}\rVert^2 \right] \\
            \vspace{-2mm}
            &= \!\! \sum_{i \in \mathcal{I}} \!\sum_{k \in \mathcal{K}_i} \!\! \sum_{m=1}^{M} \!\! [\mathbf{C}_{h,m,k}]_{1,1} \!\! + \!\! \frac{1}{M^2} \!\!\! \sum_{m=1}^{M} \!\!\! \left( \rm{tr}\{\mathbf{C}_{h,m,k} \} \!\! - \!\! [\mathbf{C}_{h,m,k}]_{1,1}\right),
        \end{aligned}
    \end{equation}
    where $[\mathbf{C}_{h,m,k}]{j,j}$ denotes the $j$-th diagonal element of $\mathbf{C}_{h,m,k}$. The estimation error represents the sum of the mean squared errors (MSE) for channel estimation across all users.
    \vspace{-3mm}
    \section{MF-RIS Beamforming Design}
	
	\subsection{Problem Formulation}
    Based on the proposed LS estimators, we can formulate the estimation error minimization problem as
    \begin{subequations} \label{origin_problem}  
    \begin{align}
        \setlength\abovedisplayskip{2pt}
        \setlength\belowdisplayskip{2pt}
        \label{origin_objective}
        \min \limits_{\bm{\varphi}_{R}, \bm{\varphi}_{T}}
        \ & \epsilon \\
    			{\rm s.t.} \
        \label{origin_RISamplify_constraint}
        \ & \lvert \varphi_{i,n} \rvert \leq \sqrt{\beta_{\max}}, \ \forall n, \forall i \in \mathcal{I} \\
        \label{origin_pattern_amplify_constraint}
        \ & \lvert \varphi_{R,n} \rvert^2 + \lvert \varphi_{T,n} \rvert^2 \leq \beta_{\max}, \ \forall n.
    \end{align}
    \end{subequations}
    	Specifically, constraints \eqref{origin_RISamplify_constraint} and \eqref{origin_pattern_amplify_constraint} represent the amplitude constraint of each MF-RIS element. 
    	{Problem \eqref{origin_problem} is challenging to solve directly.
    	Thanks to the proposed estimator is a MVU estimator, we can obtain its CRLB, which is given by $[\mathbf{C}_{h,m,k}(\mathbf{h}^i_{m,k})]_{l,l} = [\mathcal{I}^{-1}(\mathbf{h}^i_{m,k})]_{l,l}$,} 
        {where $\mathcal{I}(\mathbf{h}^i_{m,k})$ is the Fisher information for $\mathbf{h}^i_{m,k}$.}
    	Hence, the lower bound of the estimate variables is expressed as
    		\begin{equation}
    			\renewcommand{\theequation}{11}
    			\setlength\abovedisplayskip{2pt}
    			\setlength\belowdisplayskip{2pt}
    			\label{channel_estimator_CRLB_2}
    			[\mathbf{C}_{h,m,k}(\mathbf{h}^i_{m,k})]_{l,l} \geq \frac{1}{[\mathcal{I}^{-1}(\mathbf{h}^i_{m,k})]_{l,l}}, \forall l, 
    		\end{equation}
        where the equality holds when the $\mathbf{C}_{h,m,k}$ is a diagonal matrix. It is noteworthy that the objective function \eqref{origin_objective} is related to the diagonal entries of $\mathbf{C}_{h,m,k}, \forall k$. For any arbitrary $\mathbf{C}_{h,m,k}$, it can be transformed into a diagonal matrix using a unitary matrix. Inspired by \cite{2020ICASSP}, we utilize the scaled DFT matrix to simplify problem \eqref{origin_problem} into a more tractable form.
	Specifically, the product of MF-RIS beamforming and the user-MF-RIS channels is expressed as a part of each row of the scaled DFT matrix.
	Then, we design the reflection and refraction beamforming with the AO algorithm.
	This method makes the MF-RIS beamforming design effectively.
	\vspace{-5mm}
	\subsection{DFT Matrix-Based MF-RIS Beamforming Design}
	Denote $\mathbf{D}_{L, N+1}$ as the first $N+1$ columns of the $L$-point DFT matrix, and the last $N$ columns of the $\mathbf{D}_{L, N+1}$ as $\mathbf{D}_{L, N}$.
	Then, let $\bm{\varphi}_{i}^{\rm H}\mathbf{G}_{m}$ be the scaled $l$-th row of $\mathbf{D}_{L, N}$ for the $m$-th BS antenna, then we have
	\begin{equation}
		\setlength\abovedisplayskip{2pt}
		\setlength\belowdisplayskip{2pt}
		\label{origin_MFRIS_reflection_coefficient}
		[\bm{\varphi}_{i}\mathbf{G}^{\rm H}_{m}, \cdots, \bm{\varphi}_{i}\mathbf{G}^{\rm H}_{m}]^{\rm H} = a_i\mathbf{D}_{L, N},
	\end{equation}
    {where $a_i \in \mathbb{R}^{+}, i \in \mathcal{I}$ is a scaled parameter to adjust the amplification factors of the MF-RIS.}
	It is observed that $\bm{\varphi}_{i}^{\rm H}\mathbf{G}_{k}$ and $\bm{\varphi}_{i}^{\rm H}\mathbf{G}_{k^{'}}$ only has the phase difference, and we have the following equation
	\begin{equation}
		\setlength\abovedisplayskip{2pt}
		\setlength\belowdisplayskip{2pt}
		\label{the_phase_difference}
		\bm{\varphi}_{i}^{\rm H}\mathbf{G}_{k} = e^{j\Delta\theta_{k, k^{'}}}\bm{\varphi}_{i}^{\rm H}\mathbf{G}_{k^{'}},
	\end{equation}
	where $\Delta\theta_{k, k^{'}}$ is the phase difference in antenna $k$ and $k^{'}$.
	Hence the correlation matrix of channel estimators is not related to the antenna \cite{2023WCLActiveRISesti}, and we can replace the $\mathbf{G}_{m}$ as $\mathbf{G}_{1}$ in \eqref{origin_MFRIS_reflection_coefficient}. 
    Let $\mathbf{G}_{1} = \mathbf{G}$, then the beamforming at $l$-th pilot slot is obtained by
	\begin{equation}
		\label{MFRIS_coefficient}
		\bm{\varphi}_{i}^{\rm H} = a_i(\mathbf{d}_{l}^{\rm H}\mathbf{G}^{-1})^{\rm H}, l = 1, 2, \ldots, L,
	\end{equation}
	where $\mathbf{d}_{l}^{\rm H}$ is the $l$-th row of $\mathbf{D}_{L, N}$.
    Note that the DFT matrix is orthogonal matrix, hence we have $\mathbf{\Psi}_{m,i}\mathbf{\Psi}_{m,i}^{\rm H} = Na_i^2\mathbf{I}_L$,  then substitute it into \eqref{estimation_channel_correlation_matrix}, we have
        \begin{equation}
    	\label{simplify_channel_correlation_matrix}
            \mathbf{C}_{h,m,k} \!=\! \frac{N\!( a_R^2 + a_T^2)\sigma_s^2 \!+ \! \sigma^2}{P_k}(\mathbf{\Theta}_{m,i}^{\rm H}\mathbf{\Theta}_{m,i})^{-1}, \forall m.
        \end{equation} 
    Furthermore, note that $\mathbf{\Theta}_{m,i}^{\rm H}\mathbf{\Theta}_{m,i} = L\mathrm{diag} \{1, a_i^2, \cdots, a_i^2\}$, and substituting \eqref{simplify_channel_correlation_matrix} into \eqref{origin_objective}, 
    we can obtain the lower bound of the objective function \eqref{origin_objective}, which is formulated as
        \begin{equation}
    	\setlength\abovedisplayskip{2pt}
    	\setlength\belowdisplayskip{2pt}
            \label{final_objection_function}
            \epsilon(a_R, a_T) = \sum\nolimits_{i \in \mathcal{I}} \sum\nolimits_{k \in \mathcal{K}_i} \left[\epsilon_{d,k} + \epsilon_{f,k} \right],
        \end{equation}
        where $\epsilon_{d,k}$ and $\epsilon_{f,k}$ represent the estimation errors for the user-BS and user-RIS channels of user-$k$, respectively.
        $\epsilon_{d,k}$ and $\epsilon_{f,k}$ are given by
        \begin{equation}
            \begin{aligned}
            \label{estimation_variables_single_channel}
            \!\! \epsilon_{d,k} \! &= \! \frac{1}{M}\mathbb{E}\left[\lVert \mathbf{h}^i_{k} -  \hat{\mathbf{h}}^i_{k}\rVert^2 \right] \! = \!\!	\frac{N(a_R^2 \!+ \! a_T^2)\sigma_s^2 \! +\! \sigma^2}{P_kL}, \\
            \!\!\!\! \epsilon_{f,k} \! &= \! \frac{1}{N}\mathbb{E}\! \left[\lVert \mathbf{f}^i_{k} \! - \! \hat{\mathbf{f}}^i_{k}\rVert^2 \right] \! = \!
            \frac{N(N(a_j^2a_i^{-2}  + 1 )\sigma_s^2 \!\! + \!\! a_i^{-2}\!\sigma^2)}{P_kLM},
            \end{aligned}
        \end{equation}
    where $j \in \mathcal{I}$ and $i \neq j$. Then, we simplify problem \eqref{origin_problem} as
\begin{subequations} 
	\setlength\abovedisplayskip{2pt}
	\setlength\belowdisplayskip{2pt}
	\label{adjuster_problem}
	\begin{align}
		\label{adjuster_objective}
		\min \limits_{a_R, a_T  \in \mathbb{R}^{+} }
		\ & \epsilon(a_R, a_T) \\
		{\rm s.t.} \
		\label{adjuster_problem_constraint}
		\ & \eqref{origin_RISamplify_constraint}, \eqref{origin_pattern_amplify_constraint}.
	\end{align}
\end{subequations}
Problem \eqref{adjuster_problem} is still non-convex due to the coupled reflective and refractive parameters.
 \begin{algorithm}[t]
	\caption{AO-Based Algorithm for Solving Problem \eqref{adjuster_problem}}
	\label{algorithm::1}
	\begin{algorithmic}[1]
		\renewcommand{\algorithmicrequire}{\textbf{Initialize}}
		\renewcommand{\algorithmicensure}{\textbf{Output}}
		\STATE \textbf{Initialize} $ a^{(0)}_R, a^{(0)}_T$, convergence tolerance $\Delta$, the maximum iteration number $L_1$, and set $\tau=0$.
		\REPEAT
		\STATE Fixed $a_T^{(\tau)}$, calculate $a_R^{(\tau + 1)}$ by \eqref{optimal_a_r}.
		\STATE Given $a_R^{(\tau + 1)}$, calculate $a_T^{(\tau + 1)}$ by \eqref{optimal_a_t}.
		\STATE Update $\tau \leftarrow \tau + 1$.
		\UNTIL $ \left| \epsilon^{(\tau + 1)} - \epsilon^{(\tau)} \right| \leq \Delta $
		or $\tau \geq L_1$
		\STATE \textbf{Output} the converged amplification parameters $a_R^{\star}$ and $a_T^{\star}$.
	\end{algorithmic}
\end{algorithm}
To address this problem, we propose a AO-based algorithm. \par
Fixed $a_T$, the sub-problem of the reflective parameters optimization is formulated as
\begin{subequations} 
	\setlength\abovedisplayskip{2pt}
	\setlength\belowdisplayskip{2pt}
	\label{origin_reflection_problem}
	\begin{align}
		\label{reflection_coefficient_objective}
		\min \limits_{a_R \in \mathbb{R}^{+}}
		\ & \epsilon(a_R) \\
            \vspace{-4mm}
		{\rm s.t.} \
		\label{reflection_problem_constraint}
		\ & \lvert \varphi_{R,n} \rvert \leq \sqrt{\beta_{\max}}, \ \forall n, \\
		\label{MFRIS_reflection_coefficient}
		\ & \bm{\varphi}_{R}^{\rm H} = a_R(\mathbf{d}_{l}^{\rm H}\mathbf{G}^{-1})^{\rm H}.
	\end{align}
\end{subequations}	
Since we assume that the user-MF-RIS channels as the LoS channel, substituting the \eqref{MFRIS_reflection_coefficient} into \eqref{origin_pattern_amplify_constraint} with the fixed $a_T$, we can obtain the range of $a_R$, which is expressed as
\begin{equation}
	\setlength\abovedisplayskip{2pt}
	\setlength\belowdisplayskip{2pt}
	\label{a_r_range}
	0 \; \textless \; a_R \leq \sqrt{\beta_{\max}\alpha - {(a_T)^2}},
\end{equation}
where $\alpha$ denotes the path loss exponent of MF-RIS-BS channel. 
It is obvious that the objective function \eqref{reflection_coefficient_objective} is convex for the reflective parameter $a_R$.
Thus, let $\frac{\mathrm{d}\epsilon(a_R)}{\mathrm{d}a_R} = 0$, we can obtain the optimal $a_R$ as follows
    		\begin{equation}
    			\setlength\abovedisplayskip{2pt}
    			\setlength\belowdisplayskip{2pt}
    			\label{optimal_a_r}
    			\!\!\!\! a_R^{\star} \! = \! \min\{ \sqrt[4]{\!\frac{N\lvert \mathcal{K}_T \rvert a_T^2\sigma_s^2 + \sigma^2}{\sigma_s^2[M \!\! + \!\! \frac{P_R}{P_T}C_R]}}, \sqrt{\beta_{\max}\alpha \! -  \!{(a_T)^2}}\},
    		\end{equation}
    		where $P_i = \sum\nolimits_{k \in \mathcal{K}_i} (\frac{1}{P_k})^{-1}$ and $C_i = M \lvert \mathcal{K}_j \rvert + N \lvert \mathcal{K}_i \rvert a_j^{-2}$.

Given $a_R$, the refractive parameters design problem can be formulated as
\begin{subequations} \label{refraction_problem}
	\begin{align}
		\setlength\abovedisplayskip{4pt}
		\setlength\belowdisplayskip{4pt}
		\label{a_t_objective}
		\min \limits_{a_T \in \mathbb{R}^{+}}
		\ & \epsilon(a_T) \\
            \vspace{-3mm}
		{\rm s.t.} \
		\label{refraction_problem_constraint}
		\ & \lvert \varphi_{T,n} \rvert \leq \sqrt{\beta_{\max}}, \ \forall n, \\
		\label{MFRIS_refraction_coefficient}
		\ & \bm{\varphi}_{T}^{\rm H} = a_T(\mathbf{d}_{l}^{\rm H}\mathbf{G}^{-1})^{\rm H}.
	\end{align}
\end{subequations}
    Similarly, the optimal $a_T$ is expressed as
    \begin{equation}
        \setlength\abovedisplayskip{2pt}
        \setlength\belowdisplayskip{2pt}
    			\label{optimal_a_t}
        a_T^{\star} \! = \! \min\{\!\sqrt[4]{\frac{N \lvert \mathcal{K}_R \rvert a_R^2\sigma_s^2 + \sigma^2}{\sigma_s^2[M  +  \frac{P_T}{P_R}C_T]}}, \sqrt{\beta_{\max}\alpha \! -  \!{(a_R)^2}}\}.
    \end{equation}
    \par  
\begin{figure*}[!h]
	  \centering
	  \subfloat[Sum MSE versus $P_{\rm W}$.]{\includegraphics[width=2.40 in]      {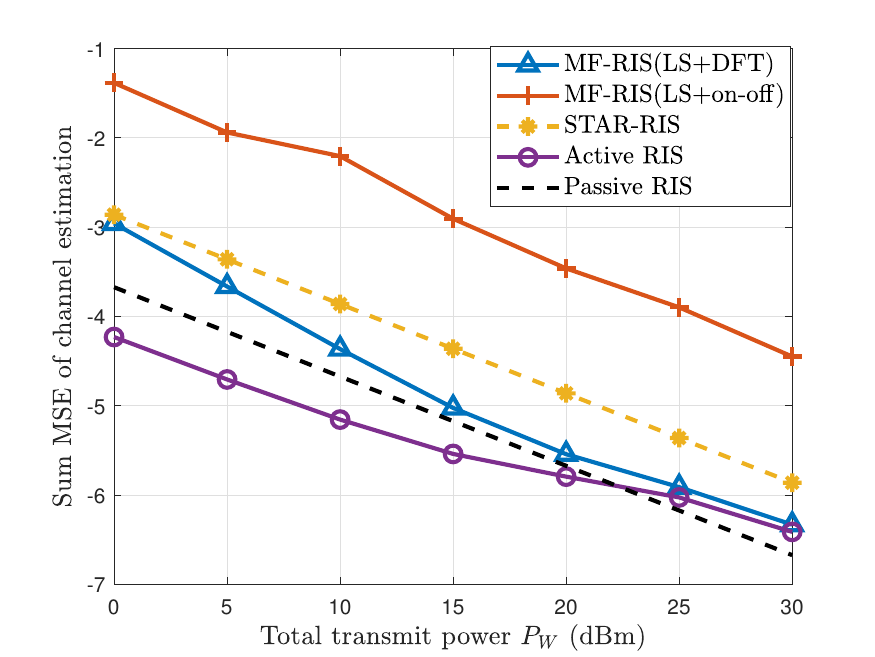}\label{change_P_2}}
        \subfloat[Sum MSE versus $d$.]{\includegraphics[width=2.40 in]{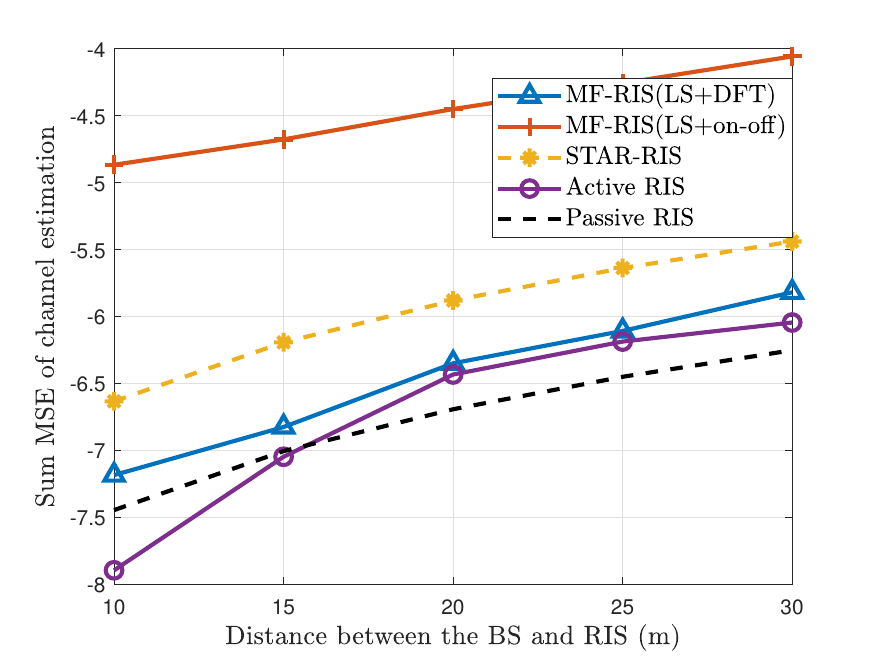}\label{change_d_2}}
        \subfloat[Sum MSE versus number of users $K$.]{\includegraphics[width=2.40 in]{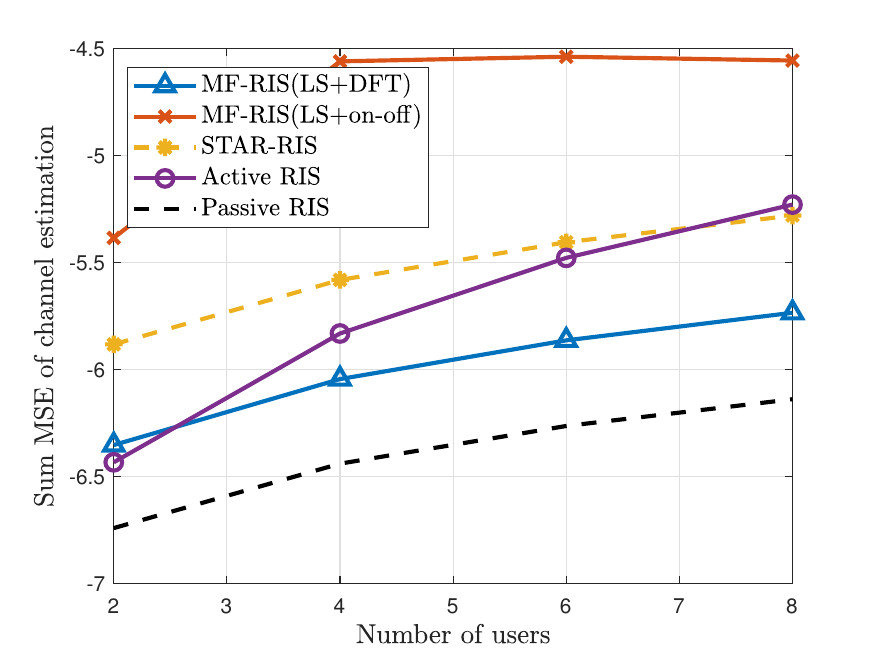}\label{multiuser_sim_2}}
        \caption{Performance comparison with existing RISs and different channel estimation methods.}
		\vspace{-5 mm}
	\end{figure*}
Finally, the MF-RIS beamforming $\bm{\varphi}_{R}$ and $\bm{\varphi}_{T}$ are obtained by substituting \eqref{optimal_a_r} and \eqref{optimal_a_t} into \eqref{MFRIS_coefficient} with the knowledge of $\mathbf{G}_{m}$, respectively.
Specifically, $\beta_{\max}$, $\alpha$, $M$, $\sigma_s^2$, and $\sigma^2$ are the necessary channel parameters to design the MF-RIS beamforming, which is the prior knowledge or can be measured precisely.
Meanwhile, the LS estimator for $\mathbf{h}^i_{m,k}$ can be simplified as
    \begin{equation}
    \label{simplify_estimator_1}
    \setlength\abovedisplayskip{2pt}
    \setlength\belowdisplayskip{2pt}
    \hat{\mathbf{h}}^i_{m,k} = \frac{1}{\sqrt{P_k}}(\mathbf{\Theta}_{m,i}^{\rm H}\mathbf{\Theta}_{m,i})^{-1}\mathbf{\Theta}_{m,i}^{\rm H}\mathbf{y}_{m,k}.
    \end{equation}
\par
\vspace{-2mm}
The amplification parameters optimization algorithm for solving problem \eqref{adjuster_problem} is given by Algorithm \ref{algorithm::1}. For the proposed AO-based MF-RIS beamforming design algorithm, 
{the optimal reflective and refractive parameters can be obtained with constant-time complexity operations.}
Hence it has a complexity of $\mathcal{O}(I_a)$, where the $I_a$ is the iteration times.
And we utilize the first $N+1$ columns of the $L$-point scaled DFT matrix to operate the $\bm{\varphi}_{i}^{\rm H}\mathbf{G}_{m}, \forall i \in \mathcal{I}$,
thus both the reflection and refraction beamforming design problems have a complexity of $\mathcal{O}(ML\log L)$ \cite{2020ICASSP}, and the total complexity of the proposed algorithm is $\mathcal{O}(2ML\log L)$. \par
Theoretically, when $a_i^{\star} = \sqrt{\beta_{\max}\alpha}$, $a_j = 0$, i.e., one of mode is unavailable or the active RIS scheme. 
In this scenario, one of region~($\mathcal{K}_r$ or $\mathcal{K}_t$) is out of RIS service, and
    \begin{equation}
    \setlength\abovedisplayskip{2pt}
    \setlength\belowdisplayskip{2pt}
    \begin{aligned}
    \label{estimation_variables_channel}
    \epsilon_{d} = \frac{N\lvert \mathcal{K}_i \rvert(a_i^{\star})^2\sigma_s^2 + K \sigma^2}{P_iL}, \epsilon_{f} = \inf.
    \end{aligned}
\end{equation}
To improve the robustness of the proposed algorithm, we assume $a_j \neq 0$ and still use Algorithm \ref{algorithm::1} to update $a_R$ and $a_T$.

\vspace{-2mm}
\section{Numerical Results}		
In the simulation, we consider a uplink system with the channel estimation for MF-RIS, where two users are randomly distributed within a circular region centered at $(0, 20, 0)$ meter~(m) with a radius of $5$~m.
The distance between the BS and the MF-RIS is $d = 20$~m. 
The reference path loss is set to {$0$} dBm, and the path loss exponents of RIS-BS is set to $2.5$, yielding {$\alpha = {10^{-3} d^{-2.5}}$} \cite{2024CLMFRISUL}. 
    Unless otherwise stated, we fix $N = 25$, $M = 8$, and $K = 2$. The maximum amplification factor of the MF-RIS is set to {$\beta_{\max} = 19$} dB.}
    In addition, the thermal noise and the receive noise are set to $\sigma_s^2 = -70$~dBm and $\sigma^2 = -80$~dBm, respectively \cite{2023WCLActiveRISesti}.
   We allocate all users as equal power, i.e., $P_k = P_{k^{'}} = 20$ dBm.
For performance comparison, we deploy the on-off state beamforming design with LS estimator in MF-RIS \cite{2020RISChannelEstimationTCOM}.
And we utilize LS estimators in conjunction with DFT-based beamforming design for the simultaneous transmitting and reflecting RIS~(STAR-RIS) \cite{2023SRARTCOM}, active RIS \cite{2021ActiveTWC}, and passive RIS \cite{2019WUTWC} schemes.
Specifically, for STAR-RIS scheme, we set the $\beta_{l,R,n} = \beta_{l,T,n} = 0.5, \forall l, n$. \par
	Fig. \eqref{change_P_2} indicates the sum MSE versus the total transmit power $P_{\rm W}$.
    Generally, as $P_{\rm W}$ increases, the sum MSE decreases in all schemes.
   Specifically, we observe that the sum MSE of the DFT-based MF-RIS beamforming design algorithm outperforms that of the on-off-based. 
    This is because the on-off scheme considers only the switching state of each MF-RIS element without optimizing channel estimation errors.
    Notably, the sum MSE of the DFT-based MF-RIS approaches that of the active RIS when $P_{\rm W} \geq 20$~dBm.
    Furthermore, we note that the sum MSE of the DFT-based MF-RIS and active RIS schemes surpasses STAR-RIS and is comparable to the passive RIS scheme.
    This is attributed to the ability of MF-RIS and active RIS to enhance the incident signal, thereby reducing channel estimation errors.
	Fig. \eqref{change_d_2} depicts the sum MSE versus the distance between RIS and BS $d$. 
	As the RIS moves further from the BS, the sum MSE increases. Both active and passive RIS schemes achieve a lower sum MSE, as they only reflect the impinging signal. It is observed that the binary-state scheme has the highest sum MSE in all schemes.
	
	
	Fig. \eqref{multiuser_sim_2} indicates the sum MSE versus the number of users. 
	It is observed that the sum MSE increases across all schemes as the number of users grows. Additionally, the binary-state MF-RIS beamforming design strategy results in a higher sum MSE compared to the DFT-based.
    Notably, when $K \geq 2$, the sum MSE of the MF-RIS scheme surpasses that of the active RIS scheme. 
    This advantage arises from the ability of MF-RIS to serve users in the refractive space, thereby reducing channel estimation errors. 
    When $K = 8$, the active RIS scheme even performs worse than the STAR-RIS scheme. 
    This further indicates that the proposed algorithm effectively mitigates the influence of thermal noise in MF-RIS.
\vspace{-4mm}
\section{Conclusion}
   In this paper, an LS-based estimator was proposed to solve the channel estimation problem in MF-RIS-aided multi-user communications. 
    Under the property of DFT matrix, we reformulated the channel estimation errors minimization problem as a more tractable form. 
    To solve the coupled MF-RIS beams designing problem with introduced thermal noise in MF-RIS channel estimation, we proposed an AO-based MF-RIS beamforming strategy that efficiently designs the reflection and refraction parameters while reaching the CRLB.
	Notably, we derived closed-form expressions for the channel estimation errors and the MF-RIS beamforming.
	Simulation results validated the efficiency of the proposed channel estimator and the proposed beamforming design algorithm.

	\bibliographystyle{IEEEtran}
	\bibliography{IEEEabrv,ref}
\end{document}